\documentstyle[sprocl,epsf,cite,wrapfig]{article}

\newif\ifpreprint
\preprinttrue

\ifpreprint
\fi

\begin{document}

\title{Factorization for Hard Exclusive Electroproduction
\ifpreprint
   \footnote{To appear in Proceedings of Semi-Exclusive Workshop, 
             Jefferson Lab, June 19--22, 1999.}
\fi
      }

\author{John C. Collins}

\address{
    Penn State University, 104 Davey Lab, \\
    University Park PA 16802, U.S.A.
}

\date{July 24, 1999}

\maketitle
\abstracts{
    This talk summarized the proof of hard-scattering
    factorization for hard exclusive electroproduction processes:
    deeply virtual Compton scattering and exclusive meson production.
}

\section{Introduction}
\label{sec:introduction}

The processes discussed in this talk were $ep \to eMp$, $ep \to
e\gamma p$, and $ep \to e \mu^+ \mu^- p$, all in the deep-inelastic
region where $Q$ is large.  The factorization theorems that I will
discuss provide a sound basis for the phenomenology of these
processes, which was discussed extensively at this workshop.  They
express the amplitudes for the processes in terms of operator matrix
elements and perturbatively calculable coefficient functions.

As always, the fundamental problem is that we do not know how to solve
QCD exactly, and we must appeal to approximations.  In high energy
processes, such as we consider here, large ranges of scales and of
rapidities are important.  The factorization theorems demonstrate that
enormous simplifications occur in suitable asymptotic limits, and then
enable calculations to be done in perturbation theory, where
asymptotic freedom implies that hard scattering coefficients etc.\ can
be usefully approximated by low-order Feynman graph calculations.

My presentation of the arguments tends to be graphical, since I find
this is the easiest way to exhibit their structure.  However they
correspond to quantitative mathematical proofs.  The principles are
quite general; the processes under discussion turn out to provide the
simplest situations for explaining how to prove factorization
theorems.  Although the proofs are apparently based on Feynman graph
arguments, some of the key elements are actually non-perturbative, as
I will emphasize.

\section{The simplest case: muon-pair production}
\label{sec:muons}

\begin{figure}
\begin{tabular}{cc}
     \epsfxsize=3.8cm \epsfbox{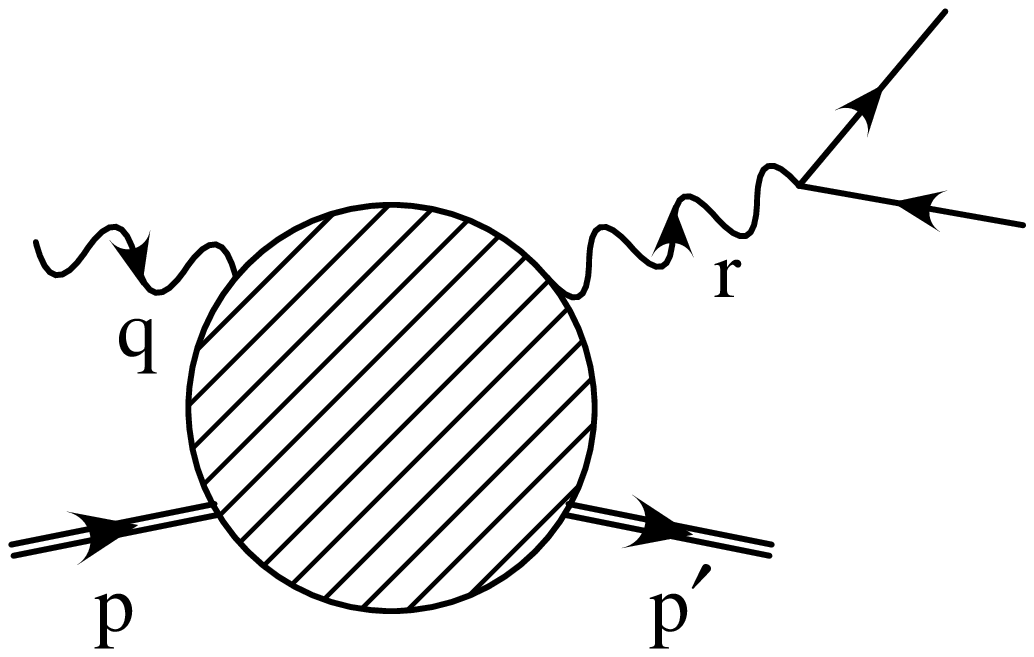}
   &
     \epsfxsize=3.8cm \epsfbox{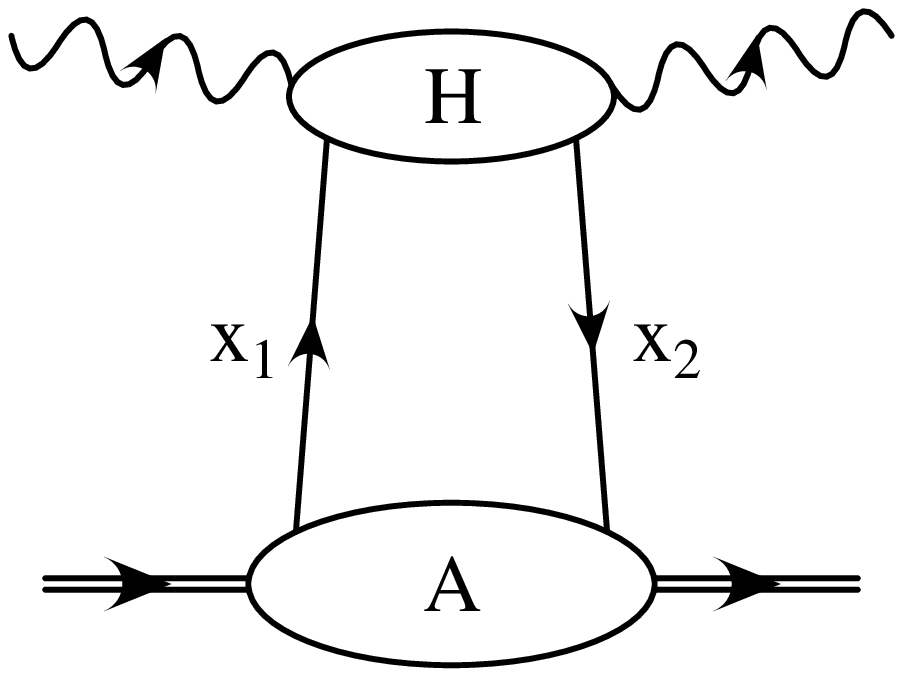}
\\
   \parbox[t]{0.45\textwidth}{
      \caption{Deeply virtual muon-pair production.}
      \label{DVMM}
   }
   &
   \parbox[t]{0.45\textwidth}{
      \caption{Leading regions for deeply virtual muon-pair 
               production.}
      \label{DVMM.lead}
   }
\end{tabular}
\end{figure}

The simplest case is the deep-inelastic production of high-mass muon
pairs: $ep \to e \mu^+ \mu^- p$, for which the kinematic variables are
defined in Fig.\ \ref{DVMM}.  The asymptotics can be treated by the
simplest version of the light-cone expansion \cite{AZ,Muller.et.al}.
However, I will explain methods that generalize to more complicated
processes, such as meson production \cite{CFS}.  Other relevant
references are \cite{Rad,Ji,CF}.

From the momenta defined in Fig.\ \ref{DVMM}, we define a set of 4
scalar variables to specify the kinematics of the process by: $Q =
\sqrt{-q^2}$, $\alpha = r^2/Q^2$, $x = Q^2 / 2p \cdot q$, and $t =
(p-p')^2$.  We consider the process at large $Q$, with $x$, $\alpha$
and $t$ fixed.  Then the leading-power part of the amplitude is given
by regions symbolized in Fig.\ \ref{DVMM.lead}.  There the lines in
the upper bubble, $H$, have large virtualities, of order $Q^2$, and
the lines in the lower bubble are approximately collinear to the
proton and have relatively low virtualities.  Power-counting arguments
\cite{CFS,Ji,Rad,CF} show that only two lines connect the two bubbles,
except that extra longitudinally polarized gluon lines can join the
two bubbles; these can be eliminated by a gauge transformation.

The methods for the power-counting can be reduced \cite{CFS} to
dimensional analysis for the finite-order graphs used for calculating
$H$, together with an analysis of the behavior of the meson and proton
subgraphs under boosts from the rest frames of the hadrons.  Thus the
methods are actually non-perturbative as regards the parts of the
amplitude that are fundamentally non-perturbative.

An expansion of $H$ in powers of small parameters (relative to $Q$)
gives the factorization theorem.  This expansion is conveniently
performed with the aid of light-front coordinates 
($V^\pm = (V^0 \pm V^z) / \sqrt2$), so that 
\begin{eqnarray}
    (p^+, p^-, {\bf p}_T) &=& (p^+, {\rm ~small}, {\bf 0}_T),
\nonumber\\
    q^\mu  &=& (-x p^+, Q^2/2xp^+, {\bf 0}_T),
\nonumber\\
    r^\mu  &=& (\alpha x p^+, Q^2/2xp^+, {\bf 0}_T) + {\rm ~small},
\nonumber\\
    p'^\mu  &=& (p^+ ( 1 - x - \alpha x) , {\rm ~small}, {\rm ~small}) .
\label{components}
\end{eqnarray}
The fractional momenta of the lines joining the two bubbles in Fig.\
\ref{DVMM.lead} are $x_1$ and $x_2$, and they obey $x_1-x_2 = x
(1+\alpha)$. The small variables in Eq.\ (\ref{components}) are small
compared with $Q$, and the variables $x$ and $\alpha$ have also gotten
redefined by a fraction of order $M^2/Q^2$ to give the formulae for
$q^\mu$ and $r^\mu$.

A factorization theorem results after one sums over all possibilities
for the graphs and after one neglects relatively small external
momentum components for $H$:
\begin{equation}
   A = \int d\xi  \sum_i H_i(Q^2, x/\xi, \alpha)
                         f_{i/p}(\xi, \xi - x(1+\alpha))
       + \mbox{non-leading power} .
\label{factorization}
\end{equation}
where the momentum fractions have been rewritten as $x_1 = \xi$ and
$x_2 = \xi - x(1+\alpha)$. The sum is over the flavors of the partons
joining the two bubbles.  

In this equation, there is implicitly the usual DGLAP scale-dependence
for the hard scattering and for the skewed parton densities.  To get a
valid perturbative calculation of the hard scattering coefficient
$H_i$ one should choose the renormalization and factorization scale
$\mu_{\overline{\rm MS}}$ to be of the order of a typical scale of
transverse momentum in $H$, i.e., of order $Q$.  The hard scattering
can be expanded in Feynman graphs for muon production from a quark or
gluon target, but the higher order terms have subtractions.  The
subtractions have two effects: they prevent double counting of
contributions to the amplitude from different orders of perturbation
theory, and they cancel the collinear divergences in the unsubtracted
graphs.

The skewed parton densities are matrix elements of suitable non-local
gauge-invariant operators, whose details are in Refs.\
\cite{Muller.et.al,Ji,Rad,CFS}.  Taking into account the dependence on
the polarization of the proton state, Ji \cite{Ji} decomposed this
parton density into scalar form factors.  There are a number of
different conventions on the kinematic variables and the normalization
for the parton densities, but the details are not important for my
purposes.

It is important that there exists an operator definition of the parton
densities: (a) it permits a precise definition of the non-perturbative
quantities involved, and (b) it gives a definite link to
non-perturbative physics.  Positivity constraints have been derived
for the parton densities in Ref.\ \cite{positivity}.

Since at fixed $Q$ and $t$, both the amplitude and the skewed parton
density depend on two scalar variables, one can in principle unfold
the skewed parton density from the amplitude, up to the usual issue of
flavor dependence.  This is unlike the case of deeply virtual Compton
scattering or meson production, where the amplitude depends on one
variable, and the unfolding is much problematic ({\em pace} Freund
\cite{unfold}). 

Finally, conformal symmetry, exactly valid at lowest order, gives
great simplification in the calculation of the DGLAP evolution kernels
for skewed parton distributions \cite{conformal}.

\section{Deeply virtual Compton Scattering}
\label{sec:DVCS}

Deeply virtual Compton scattering ($\gamma^* + p \to \gamma p$), is
intermediate between the case of high-mass muon-pair production that
was treated in the previous section and deeply inelastic meson
production to be treated in the next section, so I will not cover it
explicitly.  See Refs.\ \cite{CF,Ji,Rad}.

\section{Meson production}
\label{sec:mesons}

The case of meson production \cite{CFS}, $e+p \to e + M + p$, needs a
generalization of the methods used for muon pair production.  No
longer does a simple light-cone expansion work, and one must use the
methods summarized in Sec.\ \ref{sec:muons}.  The reason for this is
that there are two hadrons in the final state and therefore the
factorization theorem not only separates the short-distance part of
the scattering from the long-distance part, but also it gives separate
factors for each of the two hadrons.  The original discussions were
for vector mesons only ($\rho$, $\omega$, etc.), but Ref.\ \cite{CFS}
showed that the production of other mesons, like the $\pi$, also obey
a factorization theorem.

\subsection{Momentum regions}

\begin{figure}[h]
   \begin{center}
        \leavevmode
        \begin{tabular}{c@{~~~~~}c}
           \epsfxsize=4.2cm \epsfbox{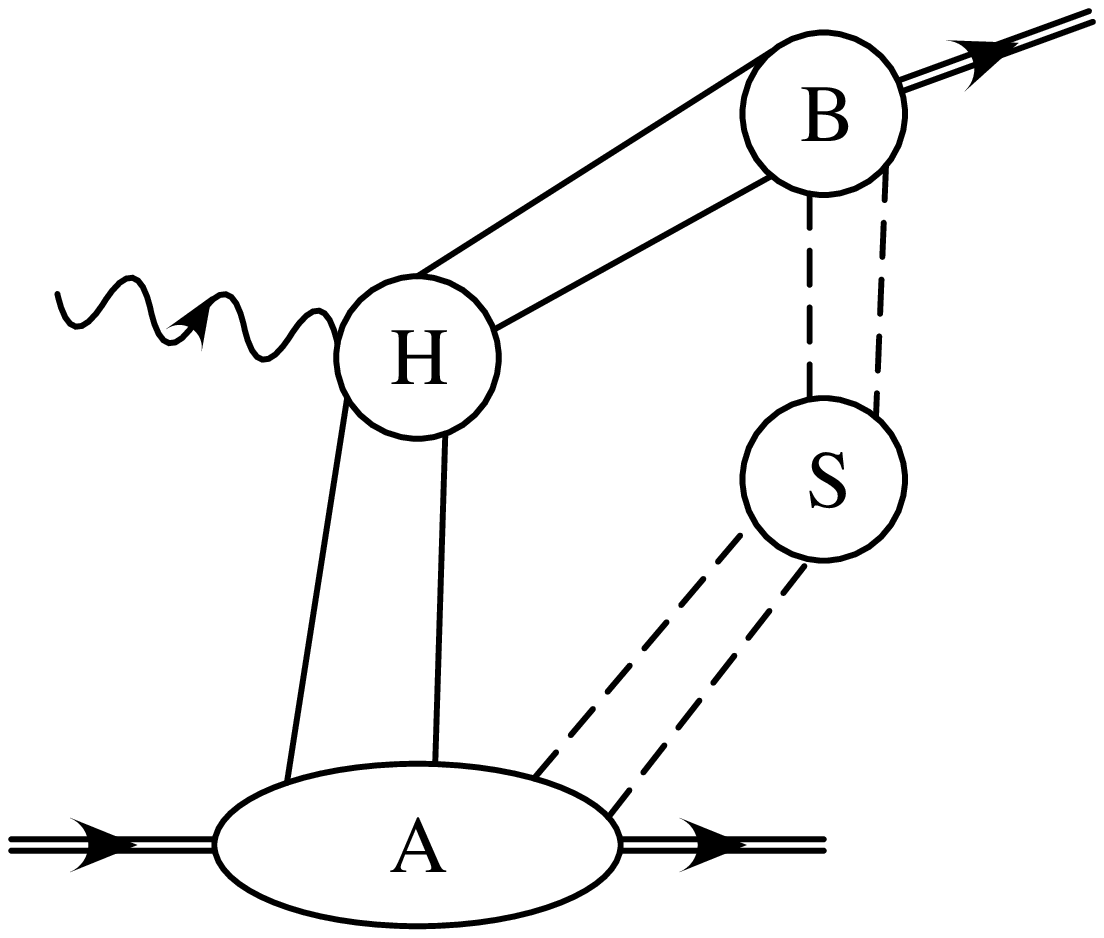} 
        & 
           \epsfxsize=4.0cm \epsfbox{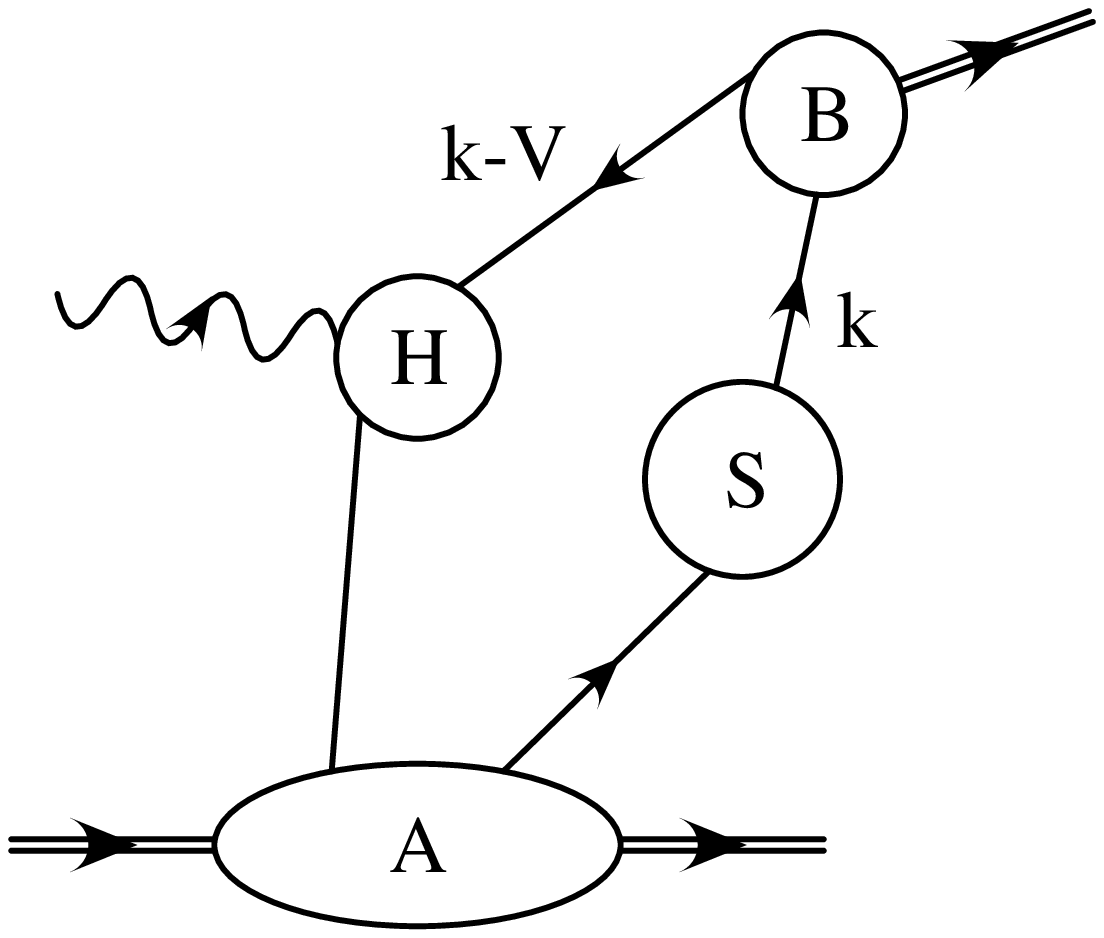}
        \\
                 (a)~~~~   &       (b)~~~
        \end{tabular}
   \end{center}
\caption{The two kinds of leading region for meson production.  
   In (a), the soft subgraph is optional.  In both graphs, there may
   be extra gluons connecting the subgraphs $A$ and $B$ to $S$ and $H$.
}
\label{fig:Reduced.Graph}
\end{figure}

The usual power-counting techniques tell us that the leading power
contributions to the amplitude come from regions symbolized by Fig.\
\ref{fig:Reduced.Graph}, which has groups of lines that are: collinear
to the target proton, collinear to the meson, soft, and hard.  The
possible soft lines joining the two collinear subgraphs are a new
feature of this process compared with muon-pair production.  Roughly,
one can characterize the sizes of the momenta in the various subgraphs
associated with the leading regions by:
\begin{eqnarray}
    H: && \left( O(Q), O(Q), O(Q) \right)
\nonumber\\
    \mbox{proton}: && \left( O(Q), O(M^2/Q), O(M) \right)
\nonumber\\
    \mbox{meson}: && \left( O(M^2/Q), O(Q), O(M) \right)
\nonumber\\
    \mbox{soft}: && \mbox{small compared with $Q$}.
\end{eqnarray}

There are two kinds of potentially leading region: 
\begin{itemize}

\item Fig.\ \ref{fig:Reduced.Graph}(a), where two lines connect each
   of the meson and proton subgraphs to the hard scattering, and only
   gluons connect the soft subgraph to the meson and proton subgraphs.

\item Fig.\ \ref{fig:Reduced.Graph}(b), where one quark connects each
   of the meson and proton subgraphs to the hard scattering, and where
   one (anti)quark plus any number of gluons connect the soft subgraph
   to the meson and proton subgraphs.

\end{itemize}
Relative to the first kind of region, the second kind gets an
enhancement because of the lower number of external lines of the hard
scattering and it gets a suppression because the external lines of the
soft subgraph are quarks. Which kind of region actually is most
important depends on the polarization of the meson.  It turns out
\cite{CFS,transverse}, by a quite non-trivial argument, that the
leading-most power only occurs for longitudinally polarized virtual
photons making either pseudo-scalar mesons or longitudinally polarized
vector mesons, and that only regions of type (a) contribute.

\begin{wrapfigure}[12]{l}{4.2cm}
   \begin{center}
     \leavevmode
     \epsfxsize=3cm \epsfbox{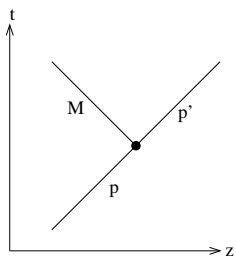}
     \caption{Space-time structure for leading regions.}
     \label{fig:space.time}
  \end{center}
\end{wrapfigure}
It is useful to observe that any of the regions involved can be
interpreted in terms of a space-time picture, Fig.\
\ref{fig:space.time}. The incoming and outgoing hadrons move in almost
light-like directions, and the hard scattering is localized close to
the intersection of the world-lines of the external hadrons.  What
will be important for our treatment of the soft subgraph in the next
section is that the lines collinear to the meson are created at the
hard scattering.  This implies that the soft interactions happen in
the final state, and not in the initial state.

\subsection{Analyticity and causality}

Consider a region of the form of Fig.\ \ref{fig:Reduced.Graph}(b).
The line coupling the hard scattering to the final-state meson has a
factor
\begin{equation}
   \frac{1}{(V-k)^2 - m^2 + i \epsilon}
   \simeq
   \frac{1}{V^2 -2 V^{-} k^{+} - k_T^2
            - 2 {\bf V}_T \cdot {\bf k}_T - m^2 + i \epsilon} ,
\label{meson.soft}
\end{equation}
where the approximation is valid to the leading power of $Q$ given
that the integration momentum $k$ is soft.  Now there are two cases to
consider for the relative sizes of the components of $k^\mu$:
\begin{itemize}

\item[(i)] The longitudinal components are small: $k^+k^- \ll k_T^2$.
   This is the region that is conventionally associated with
   rescattering corrections, i.e., small angle scattering of low mass
   objects.

\item[(ii)] The longitudinal and transverse components are comparable:
   $k^+k^- \sim k_T^2$; this is the conventional soft region.

\end{itemize}

In the first case, the only significant dependence on $k^+$ is in the
propagator (\ref{meson.soft}).  Hence the $k^{+}$ integration can be
deformed into the lower half-plane.  Then the propagator
(\ref{meson.soft}) is far off-shell, so that in effect conventional
rescattering interactions do not exist.  This derivation implements
the statement that, as in Fig.\ \ref{fig:space.time}, the meson only
has final-state interactions and that there is no causal connection
between the oppositely moving final-state particles.  The second case
has some line(s) forced off-shell, by order $QM$, where $M$ is a
typical hadron scale.

Given that case (i) is not important, a further approximation to
(\ref{meson.soft}) is valid whenever $k$ is soft:
\begin{equation}
   \frac{1}{(V-k)^2 - m^2 + i \epsilon}
   \simeq
   \frac{1}{V^2 -2 V^{-} k^{+} - m^2 + i \epsilon}
   \simeq
   \frac{1}{-2 V^{-} k^{+}+ i \epsilon} ,
\label{meson.soft1}
\end{equation}
which only depends on $k^+$, and not on $k_T$.  The second half of the
approximation depends on the non-perturbative statement that soft
lines have virtualities of at least of the order of a hadron mass.
The same approximation applies to all the soft lines coupling to the
meson subgraph in Fig.\ \ref{fig:Reduced.Graph}.  The resulting power
counting shows \cite{CFS} that of the regions in Fig.\
\ref{fig:Reduced.Graph} only (a) is leading; the soft lines
connecting to the meson are all gluons and they have a polarization
that permits certain Ward identity arguments to be used --- see Sec.\
\ref{sec:actual.leading}.  Only the first part of the approximation is
needed for the Ward identity arguments; these give the factorization
theorem.

The above arguments are quite general.  As stated earlier, the results
of the power counting encompass relevant non-perturbative effects.  In
addition, the analyticity arguments go beyond perturbation theory;
they are similar in style to the arguments of DeTar, Ellis and
Landshoff \cite{DEL}, and of Landshoff and Polkinghorne \cite{LP}.  So
we should expect that the proof is valid beyond perturbation theory,
and, in particular, as regards final-state interactions.

\subsection{``Endpoint'' contribution}

The soft contributions can also be viewed as the endpoint
contributions of the situations where lines joining the proton to the
hard scattering have small momentum fractions.  Hence the treatment of
the soft contributions has some notable implications for the
analyticity properties of the hard-scattering coefficients --- see the
papers of Radyushkin \cite{Rad} and of Collins and Freund \cite{CF}
for details.

\subsection{Actual leading contributions}
\label{sec:actual.leading}

We have now found that the actual leading-power contributions with a
non-trivial soft subgraph have the form of Fig.\
\ref{fig:Reduced.Graph}(a), where the only soft lines joining to the
meson subgraph are gluons, and where the soft momenta put lines
collinear to the meson off-shell by order $QM$.  Effectively the soft
gluons are on the same footing as the extra collinear longitudinal
gluons joining $A$ to $H$.

A Ward identity argument \cite{CFS} valid to the leading power shows
that all effects of these gluons either cancel or are effectively in
the skewed parton distributions.  This implements the statement that
the ``color of the meson is zero''.  The final result is as if the
soft subgraph $S$ is not present, and we obtain a factorization
theorem of the form:
\begin{eqnarray}
  \mbox{Amplitude} &=& H \otimes
                     \mbox{skewed pdf} \otimes
                     \mbox{meson distribution amplitude}
\nonumber\\
  && + \mbox{non-leading power}.
\end{eqnarray}
As usual, the hard scattering $H$ is perturbatively calculable in
powers of $\alpha_s(Q)$, the skewed parton distribution is the same as
in other processes, and is computed at a scale $Q$ (or of that order),
while the distribution amplitude of the meson is the same as in other
exclusive processes and is computed at a scale $Q$.

At small $x$ (e.g., for $\rho$ production at the HERA collider)
additional steps are used to discuss the Regge limit and to relate the
skewed parton density to the ordinary gluon density \cite{BFGMS}.

For $\pi$ production the skewed parton densities have a relation to
the ordinary polarized parton density, since the operators are the
same, and there are some interesting selection rules.

\section{Summary}
\label{sec:summary}

The full proof of hard-scattering factorization provides a sound basis
for the phenomenology of the processes discussed, since it follows
from the QCD Hamiltonian.  There are many non-perturbative elements in
the proof, so that it goes beyond mere perturbation theory.  This
includes definite operator definitions of the parton densities.

Part of the theorem is that amplitudes involving transverse
polarization are power suppressed.  More work is needed here to prove
an effective generalized factorization theorem for this case.

\section*{Acknowledgments}

This work was supported in part by the U.S.\ Department of Energy.


\end{document}